\documentclass[useAMS,usenatbib]{mn2e}
\usepackage{graphicx}

\title[Deep 1.4-GHz observations of  diffuse polarized emission]
      {Deep 1.4-GHz observations of  diffuse polarized emission}
\author[E. Carretti et al.]
{E.~Carretti$^{1}$\thanks{E-mail:
carretti@bo.iasf.cnr.it}, 
S.~Poppi$^{2}$,
W.~Reich$^{3}$, 
P.~Reich$^{3}$, 
E.~F\"urst$^{3}$, 
G.~Bernardi$^{1}$ \and
S.~Cortiglioni$^{1}$, C.~Sbarra$^{1}$\\
$^{1}$INAF--IASF Bologna, Via Gobetti 101, I-40129 Bologna, Italy\\
$^{2}$INAF--IRA Bologna, Via Gobetti 101, I-40129 Bologna, Italy\\
$^{3}$Max--Planck--Institut f\"ur Radioastronomie, Auf dem H\"ugel 69, 
     D-53121 Bonn, Germany}

\begin{document}

\date{Accepted xx xx xx. Received yy yy yy; in original form zz zz zz}

\pagerange{\pageref{firstpage}--\pageref{lastpage}} \pubyear{2005}

\maketitle

\label{firstpage}

\begin{abstract}
Polarized diffuse emission observations at 1.4-GHz in a
high Galactic latitude area of the northern Celestial hemisphere
are presented. The $3.2\degr\times3.2\degr$ field, centred at 
RA~=~$10^{h}$$58^{m}$, Dec~=~$+42^{\circ}$$18'$ (B1950),
has Galactic coordinates $l$~$\sim$~$172^\circ$, $b$~$\sim$~$+63^\circ$ and
is located in the region selected as northern target of the BaR-SPOrt experiment.
Observations have been performed with the Effelsberg 100-m telescope.
We find that the angular power spectra of the $E$-- and $B$--modes
have slopes of $\beta_E = -1.79\pm0.13$ and $\beta_B = -1.74\pm0.12$,
respectively.
Because of the very high Galactic latitude and the smooth emission,
a weak Faraday rotation action is expected, which allows both a fair 
extrapolation to Cosmic Microwave
Background Polarization (CMBP) frequencies and an estimate of the
contamination by Galactic synchrotron emission.
We extrapolate the $E$--mode spectrum up to 32-GHz and confirm 
the possibility to safely detect the CMBP $E$--mode signal 
in the Ka band found in another low emission region \citep{carretti05b}.
Extrapolated up to 90-GHz, the Galactic synchrotron $B$--mode looks
to compete with the cosmic signal only for models with a tensor-to-scalar
perturbation power ratio $T/S < 0.001$, which is even lower
than the $T/S$ value of 0.01 found to be accessible in the only
other high Galactic latitude area investigated to date.
This suggests that values as low as $T/S = 0.01$
might be accessed at high Galactic latitudes.
Such low emission values can allow a significant {\it red}-shift of the 
best frequency to detect the CMBP $B$--mode, also reducing the
contamination by Galactic dust, and opening interesting perspectives
to investigate Inflation models.
\end{abstract}


\section{Introduction}

The polarization of the Cosmic Microwave Background (CMB) 
is a powerful probe to investigate the early Universe.
Inflation,  scalar (density) and tensorial (gravitational waves)
primordial perturbations, and formation processes of first 
stars and galaxies can be effectively studied through
its angular power spectra
(e.g. \citealt*{zaldarriaga97,kinney99,cen03}). 
The CMB Polarization (CMBP) can be expanded in a curl-free and a
curl component called $E$-- and $B$--mode, respectively
(\citealt{zase97}; for an equivalent formalism see
\citealt*{KKS}). These fully describe the polarized emission and
have the advantage to be scalar quantities, differently from the
Stokes parameters $Q$ and $U$ that are components of a 2-spin tensor.

The emission level of the $E$--mode is a few per~cent of the CMB
temperature anisotropy signal. First detections have been recently performed
\citep{kovac02,readhead04,barkats04,leitch05,montroy05}, 
but a full characterization is far from being completed.
On subdegree scales, level and behaviour of the $E$-mode
are well predicted by the standard model, provided a number
of cosmological parameters are determined (e.g. those of the 
{\it concordance} model after the WMAP data set: \citealt{spergel03}).
A precise measurement of the power spectrum thus
represents a powerful check of the overall theoretical framework
based on both the standard model and the Inflation paradigm.

Foreground noise sources can compete with
the cosmic signal and disturb the precise measurements needed for CMBP goals.
Estimates of contaminant emissions are then crucial to define the best 
observing conditions, as the identification of the frequency window
in which the cosmic signal is accessible as a function of the sky position.
These estimates also allow the determination of the foreground properties, which,
in turn, greatly improve the effectiveness of foreground cleaning algorithms 
(e.g. \citealt{tegmark00,tucci05,verde05}).
The diffuse synchrotron emission of the Galaxy is considered
one of the most relevant polarization contaminant and is expected
to dominate the foreground contribution at frequencies
less than 100-GHz. Recent measurements performed in a 
low emission area at both 1.4 and 2.3-GHz
indicate that the synchrotron signal does not represent
an obstacle for the $E$--mode at high Galactic latitude 
and at frequencies higher than 30-GHz
\citep{bernardi03,carretti05a,carretti05b}.
However, this is just one sample that may not be representative of the
situation throughout high Galactic latitudes and
other measurements in different areas are required to confirm
this first evaluation.

The second CMBP component, the $B$--mode,
is even fainter and no information which can help
predict its level are available so far. In fact,
within the Inflation paradigm its level on degree scales 
is related to the amount of primordial tensor perturbations 
(Gravitational Waves, GW) generated
in the early whiles of the Universe by the Inflation event
(e.g. see \citealt*{ka98,kinney99,boyle05}).
Usually, this GW amount is measured by the tensor-to-scalar
perturbation power ratio $T/S$ and only upper limits are available so far
($T/S < 0.71$, 95~per~cent confidence level: \citealt{spergel03}). 
The detection of the CMB $B$--mode and the measure of its value
are thus fundamental to disentangle among different Inflation models
and to measure their relevant parameters
(e.g. \citealt{kinney99}).
Estimates of the $B$--mode emission of the foreground contaminants
(mainly synchrotron and dust emissions of the ISM)
are crucial not only to identify clean sky areas, 
but also to understand the range of Inflation models which
can be explored through CMB measurements.
The low emission area surveyed by \citet{carretti05b}
was found promising also for the detection of this CMBP component. 
In fact, these authors find that models with a GW amount down
to $T/S=0.01$ should be accessible in that area.
However, as for the $E$--mode, this is just one sample of the 
high Galactic latitude conditions and measurements
in other sky areas are needed to check whether this trend
is general.

High Galactic latitude fields are promising for CMBP investigations for
their low total intensity emission both by ISM synchrotron and dust. Among them,
\citet{carretti02} identify a couple of wide fields (about 
$30^\circ\times30^\circ$) as observing targets for the BaR-SPOrt experiment
\citep{cortiglioni03}, one for each of the northern and southern 
Celestial hemispheres.

An area in the southern field 
has been observed by \citet{bernardi03} and
\citet{carretti05b}, providing the results briefly reported above.

In this paper we present deep 1.4-GHz observations 
of an area in the northern field.
Centred at RA~=~$10^{h}$$58^{m}$,
Dec~=~$+42^{\circ}$$18'$ (B1950), the observed area  
features very high Galactic latitudes 
($l$~$\sim$~$172^\circ$, $b$~$\sim$~$+63^\circ$), which
should prevent the significant alterations
of the polarized emission  by Faraday rotation
possible at this frequency for
$|b| < 40^\circ$--$50^\circ$\citep{carretti05a}.
This leads to estimates of the $E$-- and $B$--mode power spectra
which can be safely extrapolated to the frequency range relevant
for CMB studies.

The paper is organized as follows: in Section~\ref{obsSec} we present
details of the observations along with the discussion of the obtained
maps.  In Section~\ref{specSec} we provide the analysis of polarized
angular power spectra. Finally, in Section~\ref{discSec} we discuss the
implications for CMBP measurements.

\section{Observations}\label{obsSec}

The observations were done using the Effelsberg 100-m telescope of the 
Max--Planck--Institut f\"ur Radioastronomie (MPIfR) with its 
primary-focus L-band receiver centred at 1402-MHz. 
A new 8-channel IF-polarimeter was used, where a 32-MHz wide band
was split into eight channels, 4-MHz wide each.
An additional 36-MHz wide broad band channel includes the band
covered by the $8\times 4$ MHz channels.
The addition of this multi-channel polarimeter
allows a more efficient rejection of narrow band interferences
and the derivation of the Rotation Measure (RM) of polarized emission.
The L-band receiver and its calibration was already described
by \citet{uyan98}. The receiver 
covers a wide frequency range between 1.29-GHz and 1.72-GHz, 
which, however, has the disadvantage of some cross-talk between
the Left-- and Right--Handed circular polarization band
varying with frequency. 
By observing several polarized and unpolarized
calibration sources for each observing night we found the
correction factors for each channel.
For each pixel the observed total intensity signal
was multiplied by these factors and the result was subtracted from
the corresponding $Q$ and $U$ signals.
That way residual instrumental polarization was reduced to
below 1~per~cent of the total intensity. The main calibration source
was 3C~286 assuming a flux density of  14.4-Jy and 
9.3~per~cent of linear polarization at a polarization
angle of $33\degr$.
The angular resolution at 1402-MHz was found to be 9.35-arcmin
and the conversion from the Jy~beam$^{-1}$-scale into the
main-beam-temperature-scale
is $T_{b}/S$~=~2.12~K~Jy$^{-1}$ \citep{uyan98}. 

In a first step we observed a field of 
$10\degr \times 10\degr$ centred at 
RA~=~$11^{h}$$03^{m}$, Dec~=~$44\degr$$44\arcmin$ (B1950)
 in April~2003. 
Two coverages were made by scanning the field along orthogonal directions. 
The total integration time per 4-arcmin pixel was 2-s. This map was checked
for a suitable sub-field with a largely uniformly distributed
polarized emission as indicative of a small amount of foreground Faraday
rotation effects. 
A $3.2\degr \times 3.2\degr$ field, centred at 
RA~=~$10^{h}$$58^{m}$, Dec~=~$42\degr$$18\arcmin$ (B1950) 
-- Galactic coordinates $l$~$\sim$~$172^\circ$,~$b$~$\sim$~$+63^\circ$ --
was selected for deep observations. The integration time per pixel
was always 2-s for each coverage.
The strong polarized source 3C~247 within the field 
was used to check the quality
of the individual maps, e.g. their scale and pointing accuracy.
The observations were completed end of May~2003.
The standard data reduction scheme for Effelsberg continuum
and polarization observations was applied including baseline
corrections of the individual maps by using the unsharp
masking procedure described by \citet{sofue79}.
Five coverages were used (for a total 10-s integration time per pixel),
where the data from the wide broad band channel have been used.
The five maps were subsequently combined using the {\it PLAIT} 
algorithm as described by \citet{emerson88},
which effectively destripes the maps.
The total intensity noise level is set by the confusion limit
of about 7 mJy~beam$^{-1}$ or 15-mK of brightness temperature~$T_{b}$ 
\citep{uyan99} of the Effelsberg 100-m telescope. 
The $Q$ and $U$ maps are not confusion limited for an integration time
of 10-s.  We obtain rms-noise values of about 2-mK~$T\rm _{b}$ per pixel. 
The main observation features are summarized in Table~\ref{featTab}.
%%%%%%%%%%%%%%%%%%%%%%%%%%%%%%%%%%%%%%%%%%%%%%%%%%
\begin{table}
 \centering
  \caption{Main features of the observations conducted at Effelsberg.}
  \begin{tabular}{@{}lr@{}}
  \hline
  Central frequency                  &  1402-MHz \\
  Effective bandwidth                & $36$-MHz \\
  FWHM                               & $9.35$-arcmin \\
  RA (B1950)                         & $10^{h}$$58^{m}$ \\
  Dec (B1950)                        & $42\degr$$18'$ \\
  Area size                          & $3.2^{\circ}\times 3.2^{\circ}$ \\
  Pixel size                         & $4\arcmin \times 4\arcmin$ \\
  Observation period                 & May 2003 \\
  $Q$, $U$ pixel sensitivity (flux)  & 0.9 mJy beam$^{-1}$ \\
  $Q$, $U$ pixel sensitivity ($T_b$) & 2.0-mK \\
  Gain $T_{\rm b}$/S                 & 2.12 K Jy$^{-1}$ \\
  \hline
  \end{tabular}
 \label{featTab}
\end{table}
%%%%%%%%%%%%%%%%%%%%%%%%%%%%%%%%%%%%%%%%%%%%%%%%%%
%%%%%%%%%%%%%%%%%%%%%%%%%%%%%%%%%%%%%%%%%%%%%%%%%%%%%%%%%%%%%%%%%%%%%%%%%%%%%
\begin{figure*}
  \includegraphics[angle=0, width=0.49\hsize]{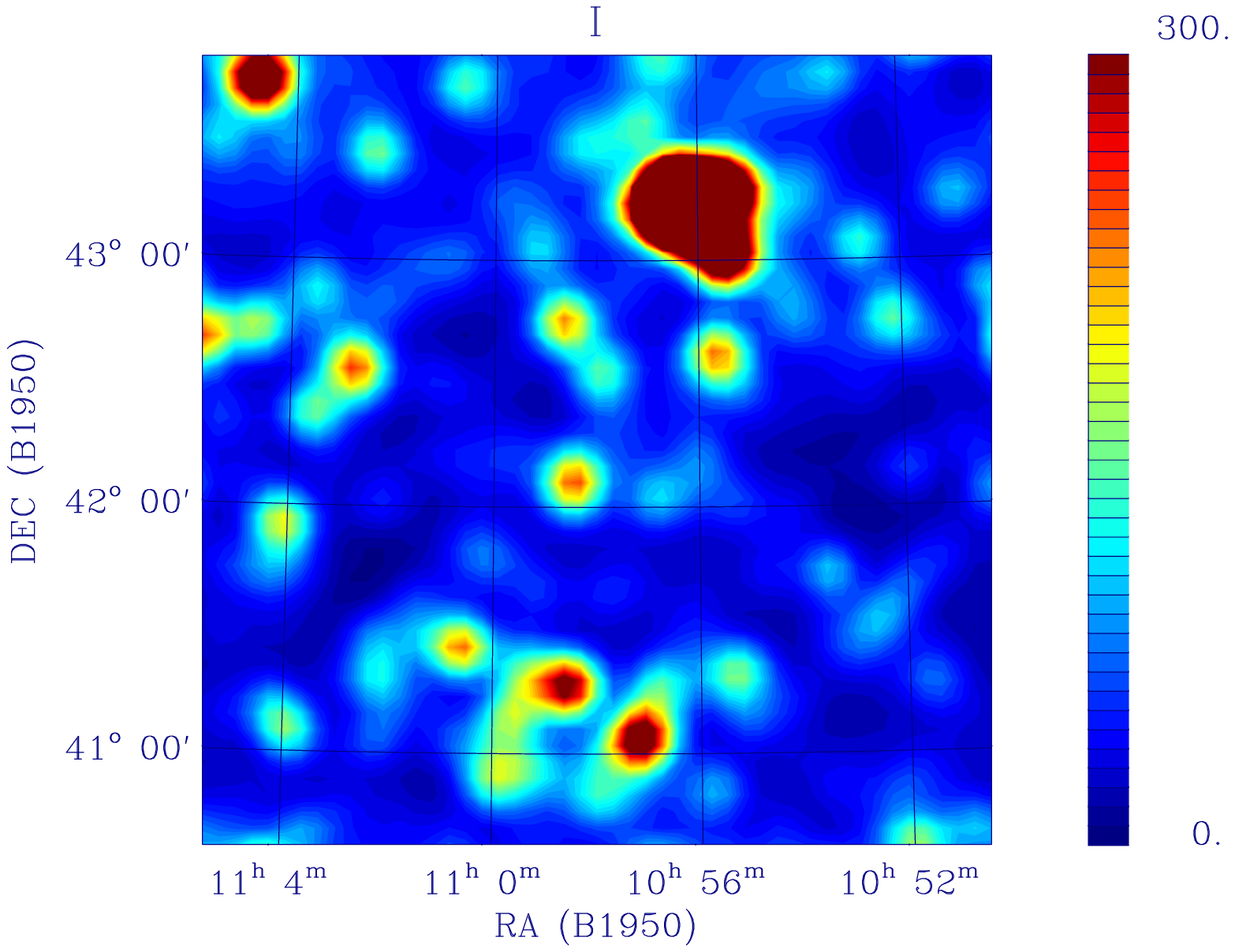}
  \includegraphics[angle=0, width=0.49\hsize]{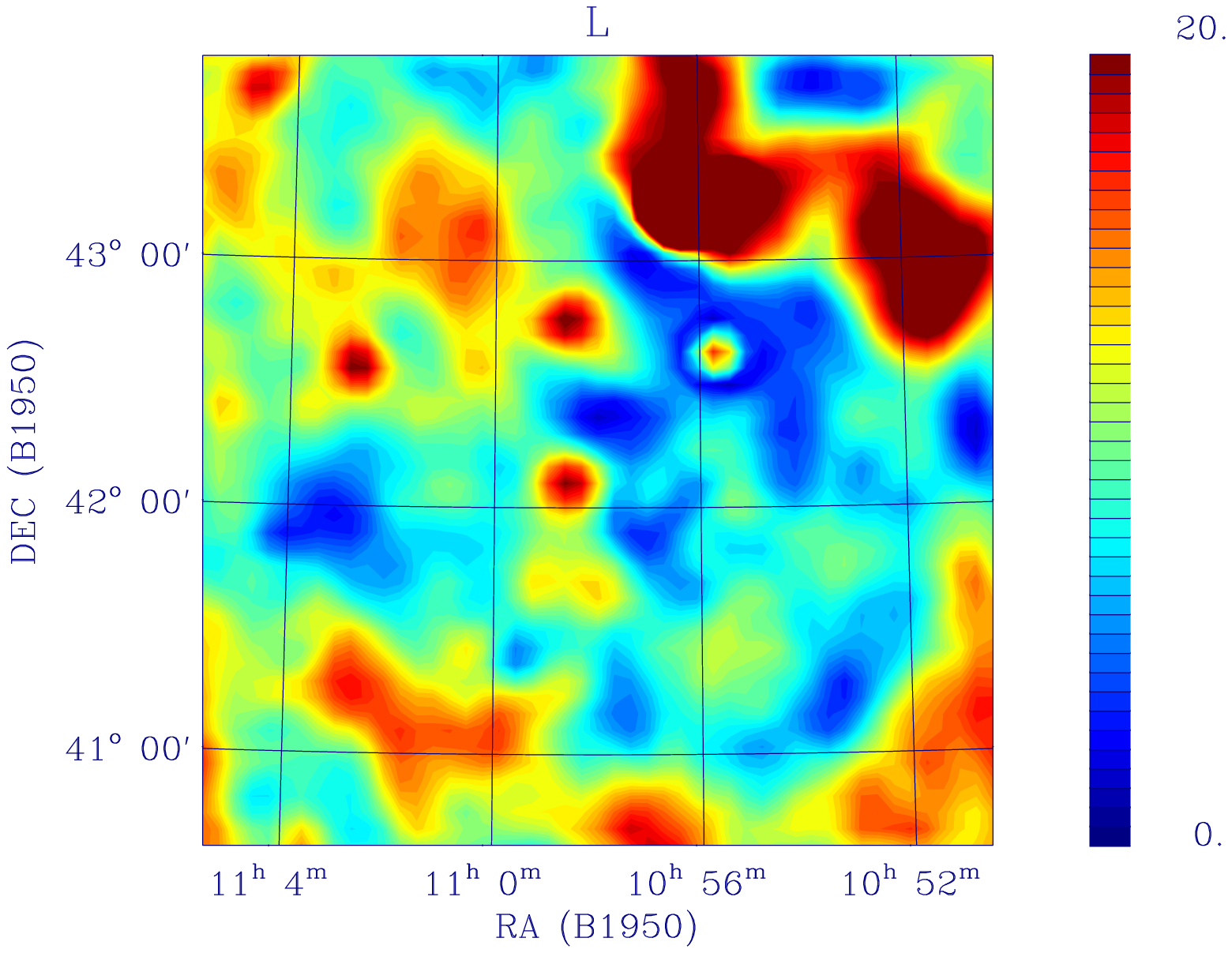}
  \includegraphics[angle=0, width=0.49\hsize]{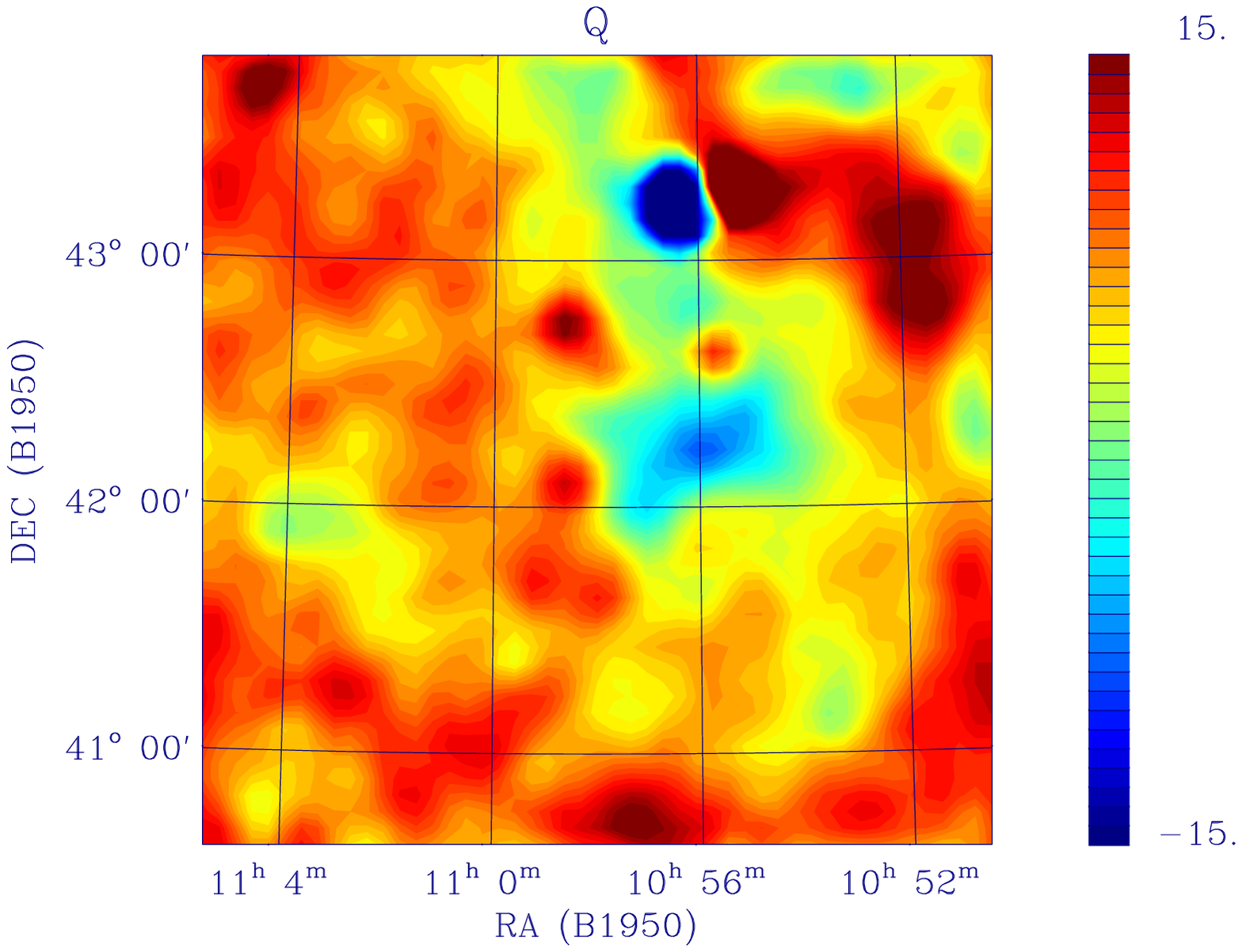}
  \includegraphics[angle=0, width=0.49\hsize]{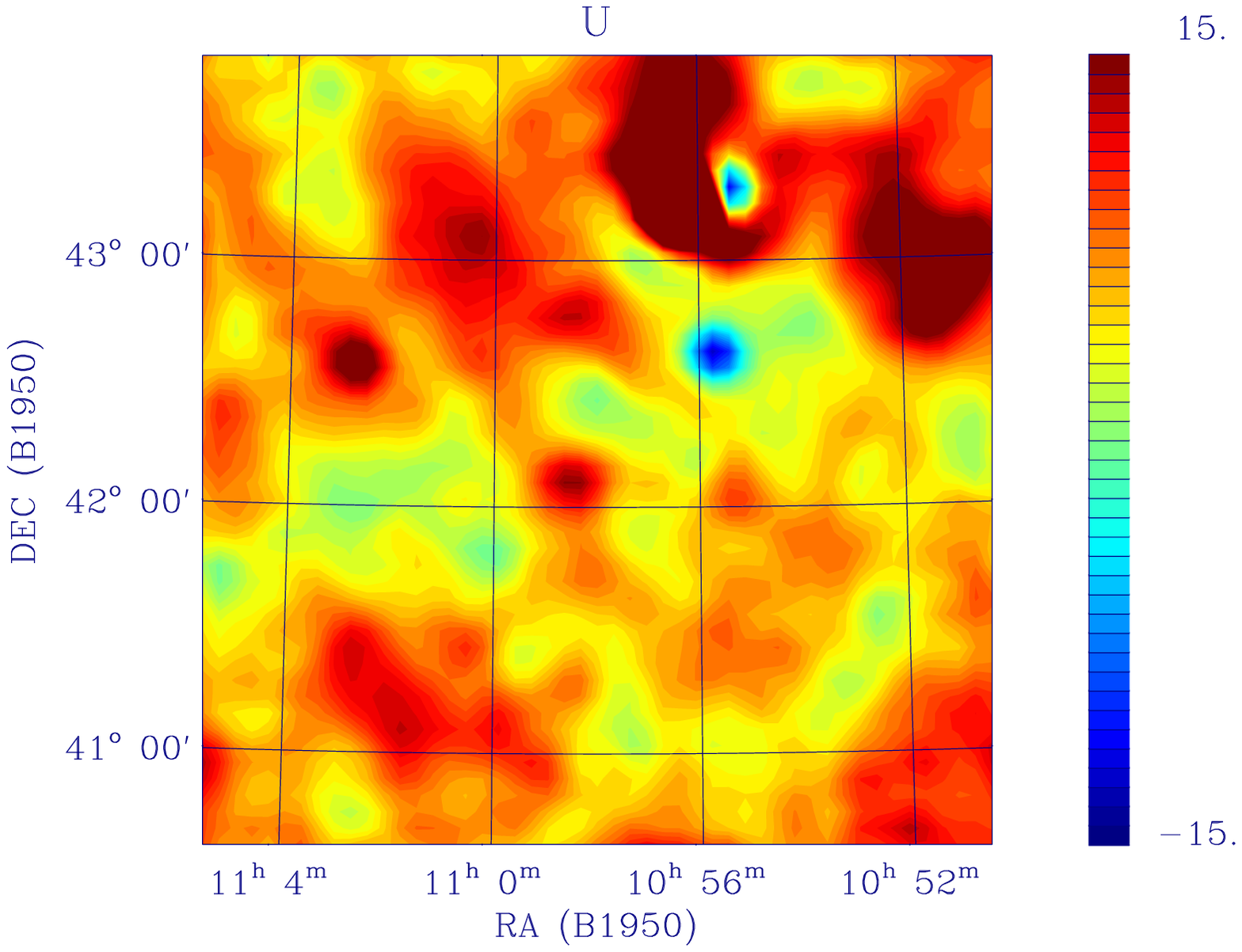}
\caption{Maps of the Stokes parameters $I$ (top-left), 
         $Q$ (bottom-left) and $U$ (bottom-right) of the observed area
         at 1.4-GHz. The polarized intensity $L=\sqrt{Q^2+U^2}$ is shown as well
         (top-right). The strongest source (3C~247) is saturated to allow
         a better view of the rest of the emission in the area. 
         Units are mK.\label{mapFig}
}
\end{figure*}
%%%%%%%%%%%%%%%%%%%%%%%%%%%%%%%%%%%%%%%%%%%%%%%%%%%%%%%%%%%%%%%%%%%%%%%%%%%%%

The Stokes $I$, $Q$, $U$ and linearly polarized intensity ($L=\sqrt{Q^2 + U^2}$)
images are displayed in Figure~\ref{mapFig}, where they have been smoothed 
to a Full Width at Half Maximum (FWHM) of 11.0-arcmin.

The field is dominated by the strong source 3C~247 both
in total and polarized intensity 
(2.9-Jy of total flux, 4.9~per~cent of polarization fraction and
$67\degr$ of polarization angle).
The rest of the $I$ image is dominated by point sources
and no evidence of diffuse emission is found on the scales
accessible to these observations.

\begin{figure}
\centering
\includegraphics[angle=0, width=1.0\hsize]{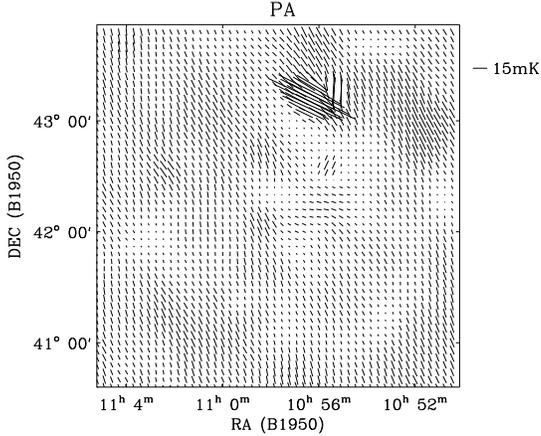}
\caption{Polarization angle map. The vector length is proportional to the 
         polarized intensity.\label{paFig}
}
\end{figure}

The polarization images are dominated by diffuse emission
and just few point sources are evident. 
Structures are present on all angular scales up to the field size,
like the feature extending close to the southern border
of the field.
Excluding the evident point sources, the polarized intensity
peaks at about 25-mK and features an rms fluctuation
of about 6-mK.
The polarization angle pattern has a regular behaviour 
(Figure~\ref{paFig}).
%%%%%%%%%%%%%%%%%%%%%%%%%%%%%%%%%%%%%%%%%%%%%%%%%%%%%%%%%%%%%%%%%%%%%%%%%%%%%
%%%%%%%%%%%%%%%%%%%%%%%%%%%%%%%%%%%%%%%%%%%%%%%%%%%%%%%%%%%%%%%%%%%%%%%%%%%%%
No peculiar {\it canal}-like feature are visible in our images, as frequently seen
in Galactic plane survey maps (as long filaments with polarized intensity
near zero, e.g. see \citealt{uyan99,haverkorn04}).
An estimate of the RM
throughout the field can be attempted thanks to our multi--band data.
However, even smoothing the signal on a 15-arcmin scale to improve the
signal-to-noise ratio,
we achieve a mean rms-error of 50~rad~m$^{-2}$. At this limit, we do not
find any significant result, in agreement with 
the values of 10--20~rad~m$^{-2}$ expected at these Galactic latitudes (e.g. see \citealt{han97}).

A direct comparison with the low
emission area observed in the southern hemisphere at the same frequency
\citep{bernardi03} is difficult because of the different angular scales
covered.
Performed with the Australia Telescope Compact Array (ATCA),
the observations of these authors have full sensitivity in
the 3--15~arcmin range, which only allows a small overlap. A more
appropriate analysis is done in Section~\ref{specSec}
through angular power spectra, which can compare specific angular scales.

Finally, polarized point sources have been subtracted. 
To identify them, we fit the maxima/minima in 
$Q$ and $U$ maps with a 2D-Gaussian beam and a constant value. 
The latter is adopted to account for the background emission.

\section{Power Spectrum Analysis}
\label{specSec}

The angular behaviour of polarized emission can be studied
using the angular power spectra of $E$-- and $B$--modes.
Besides fully describing the statistical properties
of the polarized emission with the additional advantage
to be scalar, these spectra are the quantities predicted
by the cosmological models and allow a direct comparison
with the expected cosmic signal.

\begin{figure}
\centering
\includegraphics[angle=0, width=1.0\hsize]{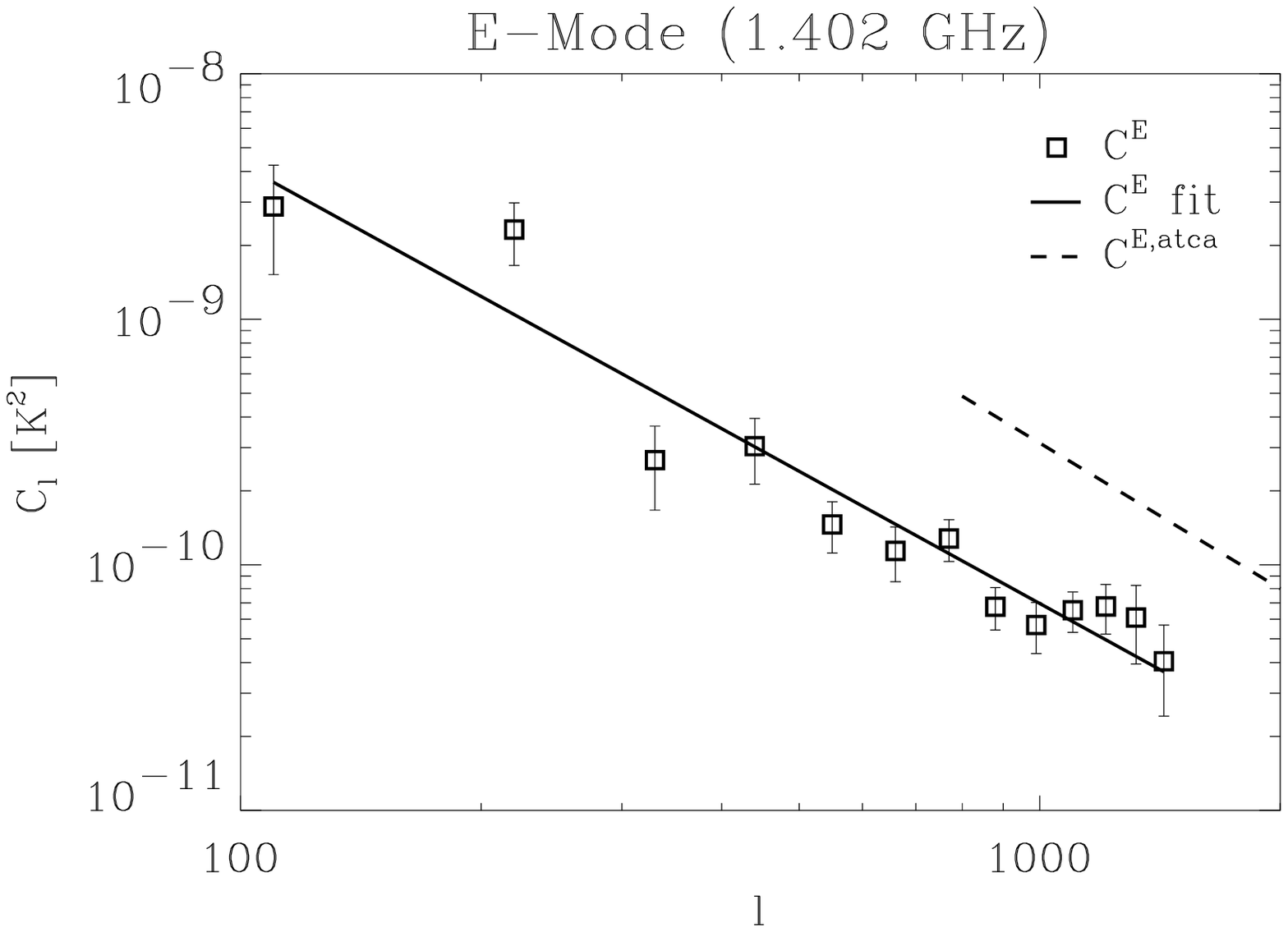}
\includegraphics[angle=0, width=1.0\hsize]{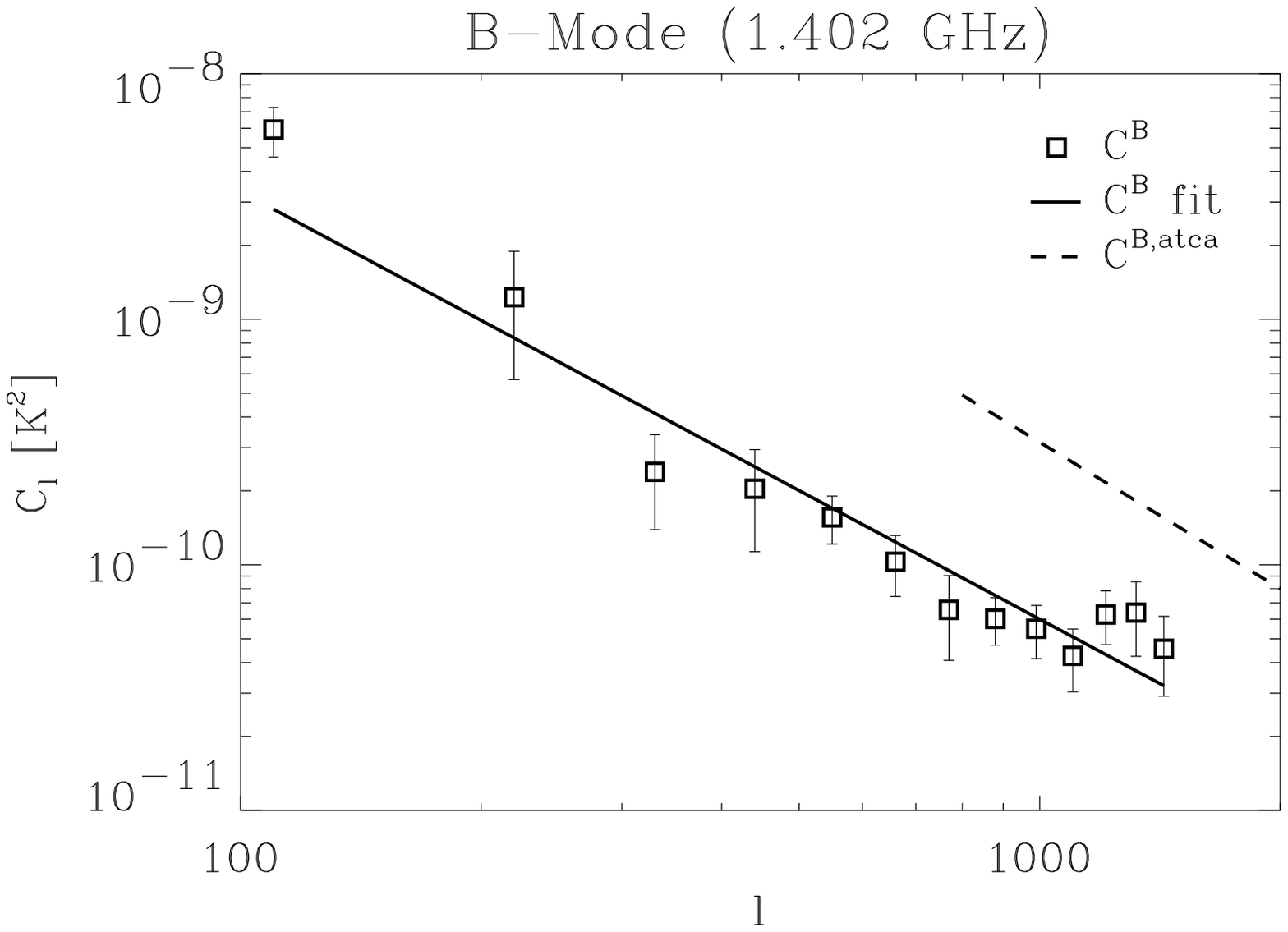}
\caption{Angular power spectra $C^E_\ell$ (top)
and $C^B_\ell$ (bottom) of the 1.4-GHz emission with their
power law fits (solid). Power spectra measured in the southern low
emission area observed with ATCA at the same frequency
are reported for comparison (dashed).
\label{specFig}}
\end{figure}

To study the properties of the diffuse synchrotron emission
we use the images cleaned from point sources.
The spectra are computed by using the Fourier method of
\citet{seljak97} and the results are shown in Figure~\ref{specFig}.
The correction by the beam window function allows us to measure the spectra
up to $\ell\sim 1400$, a value slightly larger than the multipole
corresponding to the FWHM ($\ell \sim 1150$).
As typical of the polarized synchrotron emission,
both the $E$-- and $B$--mode spectrum are well approximated
by power laws. Best fits to equation
\begin{equation}
    C_{\ell}^X = C_{500}^{X} \left({\ell \over 500}\right)^{\beta_X},\;\;
    {\rm with\;\;} X=E,\,B,
\end{equation}
as functions of the multipole $\ell$\footnote{The multipole $\ell$ is 
related to the angular scale $\theta$ by the relation $\theta$~$\sim$~$180^\circ / \ell$}
are reported in Figure~\ref{specFig} and Table~\ref{powFitTab}.
%%%%%%%%%%%%%%%%%%%%%%%%%%%%%%%%%%%%%%%%%%%%%%%%%%%%%%%%%%%%%%%%%%%%%%%%
%%%%%%%%%%%%%%%%%%%%%%%%%%%%%%%%%%%%%%%%%%%%%%%%%%%%%%%%%%%%%%%%%%%%%%%%
%%%%%%%%%%%%%%%%%%%%%%%%%%%%%%%%%%%%%%%%%%%%%%%%%%
\begin{table}
 \centering
  \caption{Fit parameters for $E$ and $B$ spectra.\label{specTab}}
  \begin{tabular}{@{}lcc@{}}
  \hline
   Spectrum     &   $C_{500}^X$ [$10^{-12}$~K$^2$]     &   $\beta_X$        \\
  \hline
  $C^E_\ell$ &  $241 \pm 21$  &  $-1.79 \pm 0.13$ \\
  $C^B_\ell$ &  $201 \pm 16$  &  $-1.74 \pm 0.12$ \\
  \hline
  \end{tabular}
 \label{powFitTab}
\end{table}
%%%%%%%%%%%%%%%%%%%%%%%%%%%%%%%%%%%%%%%%%%%%%%%%%%
The slopes $\beta_E = -1.79 \pm 0.13$ and $\beta_B = -1.74 \pm 0.12$ 
are compatible with the value of $\beta_X = -1.6 \pm 0.2$ ($X = E,B$) typical of
other sky areas when the Faraday rotation does not strongly
modify the polarized emission structure
\citep{bruscoli02,carretti05a}.

A comparison with the spectra $C^{X,{\rm atca}}_{\ell}$
of the southern deep field observed with the ATCA at the same frequency
\citep{bernardi03,carretti05a} can be done in the 
800--1400~$\ell$--range, 
which is common to the two observations 
(Figure~\ref{specFig}). 
We find that the emission of the northern area is fainter
by a factor $C^{X,{\rm atca}}_\ell/C^{X}_\ell \sim 5$.

However, this comparison could not be reliable
because of Faraday rotation effects.
\citet{carretti05a} find that a power transfer from large to 
small angular scales occurs,
as the Faraday rotation is strongly modifying
the emission pattern. This results both in a steeper slope and power enhancement 
of the power spectrum on subdegree scales.
In addition, these authors find that a transition region separates
latitudes with strong and negligible Faraday rotation effects at 1.4-GHz.
In fact, modifications of the polarized emission look negligible
at Galactic latitude $|b|> 40\degr$--$50\degr$. This is also supported by
the new 1.4~GHz maps of \citet{wolleben05}, who imaged the northern sky with
an angular resolution of 36~arcmin and a pixel sensitivity of 12~mK. 
Their maps show the signal is almost depolarized at $|b| < 30\degr$--$40\degr$
with just a patchy structure on scales smaller than few degrees.
Beyond a transition region the emission looks smooth
at $|b|>40\degr$--$50\degr$, again suggesting a negligible Faraday rotation action
above these latitudes.

The application of these considerations to the Effelsberg area 
suggests the spectra of this patch are not significantly modified.
In fact, the flat slopes $\beta_E$ and $\beta_B$ found here
indicate no relevant action by Faraday effects, also supported by
the Galactic location at $b$~$\sim$~$63\degr$,
that is above the transition region at $|b|=40\degr$--$50\degr$.
The regular pattern of the position angles 
further confirms this framework.

\citet{carretti05b} find a different situation in the southern area. 
After comparison with 2.3~GHz data taken in the same area, the 1.4~GHz power 
spectra are found to be modified. For instance, they show power enhancement
and slope steepening (even though the latter is marginal).

The enhancement of the southern area could explain at least
part of the power gap between the two fields, and makes
unreliable a direct comparison at this frequency.

\begin{figure}
\centering
\includegraphics[angle=0, width=1.0\hsize]{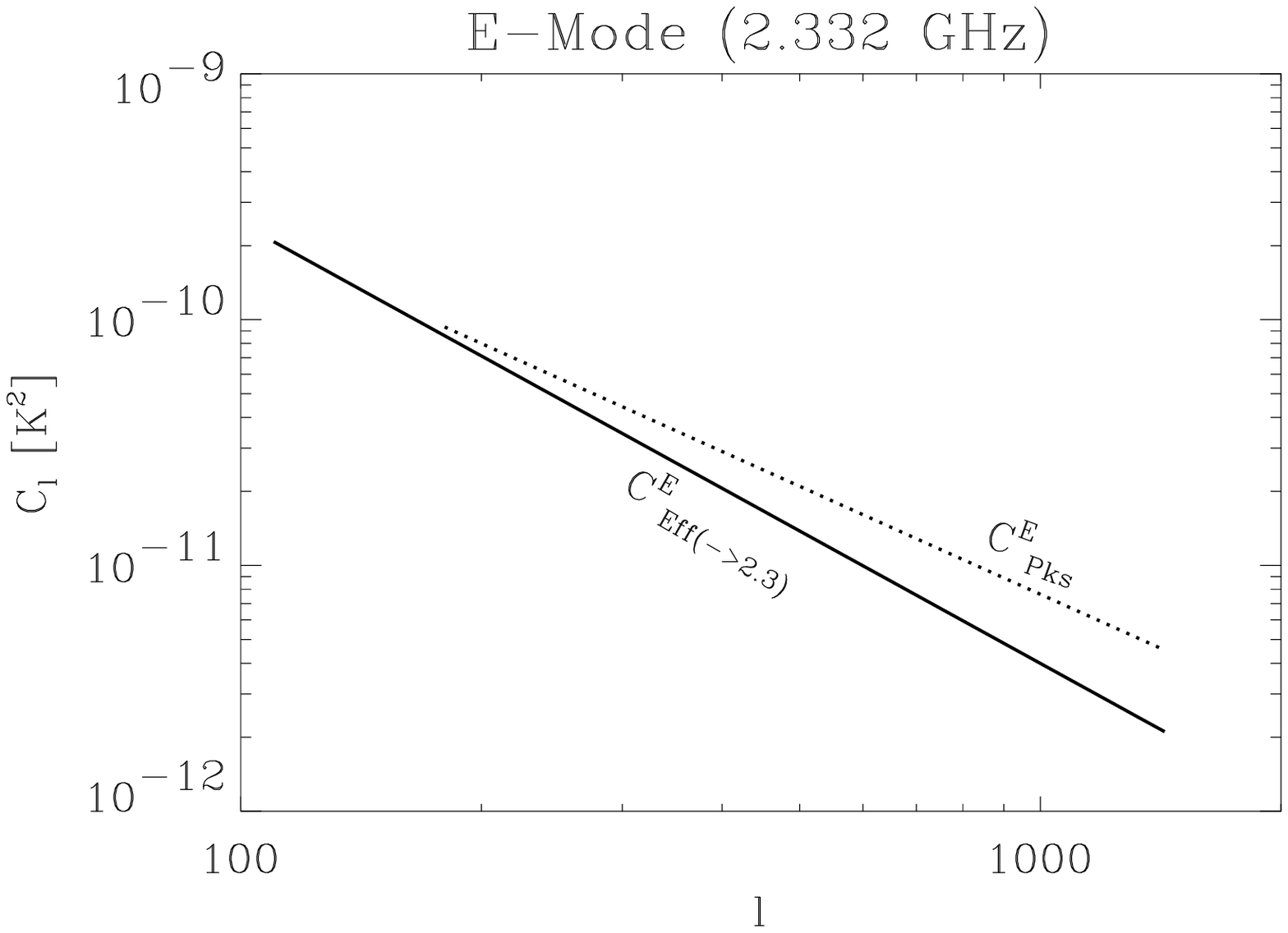}
\includegraphics[angle=0, width=1.0\hsize]{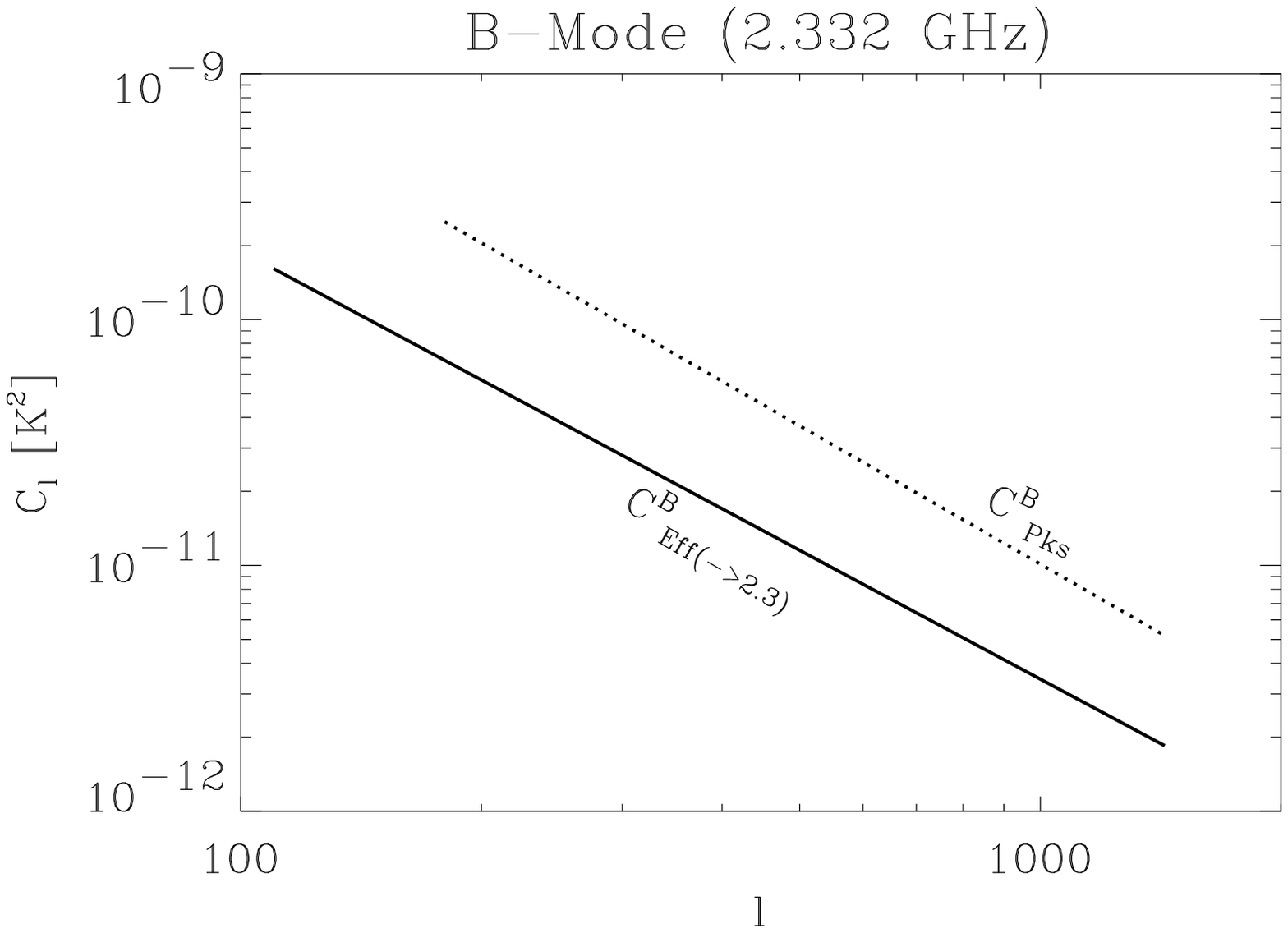}
\caption{Fits of the power spectra $C^E_{\rm Pks}$ (top) and $C^B_{\rm Pks}$ 
         (bottom)
         of the 2.332-GHz polarized emission observed in the southern area 
         with the Parkes radio telescope
         \citep{carretti05b} together with those of the northern area extrapolated
         to the same frequency ($C^E_{\rm Eff(->2.3)}$, $C^B_{\rm Eff(->2.3)})$.
\label{effPksFig}}
\end{figure}

A more reliable comparison can be done
using the 2.3-GHz emission of the southern area \citep{carretti05b},
although with the uncertainty due to the different
frequencies. 
In addition, the 2.3-GHz Parkes data 
(single-dish observations of a $2\degr\times2\degr$ patch with a 
FWHM of~8.8-arcmin) cover an $\ell$-range similar to the Effelsberg one.
The comparison is shown in Figure~\ref{effPksFig} where the spectra
of Table~\ref{powFitTab} are scaled up to the 2.332-GHz
frequency of the Parkes observations assuming an $\alpha = -2.8$ spectral
slope, as typical of the synchrotron emission at such frequencies
\citep{platania98}.
%%%%%%%%%%%%%%%%%%%%%%%%%%%%%%%%%%%%%%%%%%%%%%%%%%%%%%%%%%%%%%%%%%%%%%%
%%%%%%%%%%%%%%%%%%%%%%%%%%%%%%%%%%%%%%%%%%%%%%%%%%%%%%%%%%%%%%%%%%%%%%%%%
Differences are still present on smallest angular scales in the $E$--mode,
although the two spectra converge to similar values
on largest scales, where they look very similar. 
The $B$--mode emission observed with the Effelsberg telescope, instead,
still remains fainter than that of the southern area throughout the
$\ell$-range surveyed by the two observations.
The northern region thus results even more
promising than the southern one to search for the CMBP signal,
especially the $B$--mode.

\section{discussion}
\label{discSec}

In Section~\ref{specSec} we find indications that the Effelsberg area should
not be affected by significant
Faraday rotation effects.
Therefore, it is likely that the power spectra we find
are not significantly modified and can be safely scaled up to the CMBP
frequency range. It is worth noting that, in any case, Faraday effects
would enhance the power on these angular scales.  Therefore, 
our results would be an upper limit of the real synchrotron emission,
providing us with a worst case analysis of the contamination of the CMB
by this foreground.

Extrapolations have been performed assuming the brightness temperature 
frequency slope $\alpha = -3.1$ typical of the synchrotron
emission in the 1.4--23~GHz range \citep{bernardi04}. 
Considering that \citet{bennett03} find a continuous steepening 
of this slope from K (23-GHz) to W band (94-GHz), our assumption
should provide worst case estimates when applied to 
the 30--90~GHz range.
Brightness temperatures are converted into CMB thermodynamic 
temperatures by the factor
\begin{equation}
   c = \left[{2\,\sinh(x/2) \over x}\right]^2, \hskip 1cm x = h\nu / kT_{\rm cmb}
\end{equation}
where $\nu$ is the frequency and $T_{\rm cmb} = 2.725$~K \citep{mather99}.

\begin{figure}
\centering
\includegraphics[angle=0, width=1.0\hsize]{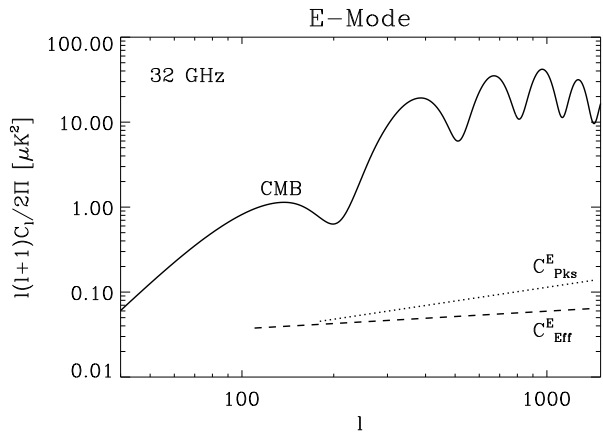}
\caption{$E$--mode angular power spectrum $C^E_{\rm Eff}$
of the 1.4-GHz data extrapolated to 32-GHz. 
A frequency spectral slope of $\alpha = -3.1$ is assumed.
The extrapolation of the 2.3-GHz spectrum obtained in the southern
area is also reported ($C^E_{\rm Pks}$).
The power spectrum expected for the CMB is shown for comparison (solid).
The cosmological parameters used are those of the concordance model
as determined in \citet{spergel03}.
\label{specCMB_E_Fig}}
\end{figure}

Figure~\ref{specCMB_E_Fig} reports the $E$--mode spectrum extrapolated
to 32-GHz, which is the lower frequency of the BaR-SPOrt experiment and
at the low end of the frequency range relevant for CMBP aims
[both the DASI \citep{leitch05} and CBI experiment \citep{readhead04} 
observed in such a band].
The synchrotron contribution is well below the cosmic component throughout
the peak region ($\ell = 150$--1500). Even the low out-of-horizon 
feature at $\ell$~$\sim$~$150$ looks well accessible. 

A comparison with the
southern region is done using the 2.3-GHz data taken in that area, more
reliable for extrapolation than the 1.4-GHz ones. 
The contamination in the two areas look similar, apart from lower values
of the northern one on smallest scales, so that our results substantially
confirm the conclusions of \citet{carretti05b}.
This result is also supported by the analyses of both the DASI and CBI experiment teams,
who find indications
that, in other two regions, the emission of this foreground is not dominant
at 30-GHz. This scenario suggests the detectability of CMBP $E$--mode
is achievable at high Galactic latitudes even in the Ka-band.

%%%%%%%%%%%%%%%%%%%%%%%%%%%%%%%%%%%%%%%%%%%%%%%%%%%%%%%%%%%%%%%%%%%%%%%%
%%%%%%%%%%%%%%%%%%%%%%%%%%%%%%%%%%%%%%%%%%%%%%%%%%%%%%%%%%%%%%%%%%%%%%%%
%%%%%%%%%%%%%%%%%%%%%%%%%%%%%%%%%%%%%%%%%%%%%%%%%%%%%%%%%%%%%%%%%%%%%%%%
\begin{figure}
\centering
\includegraphics[angle=0, width=1.0\hsize]{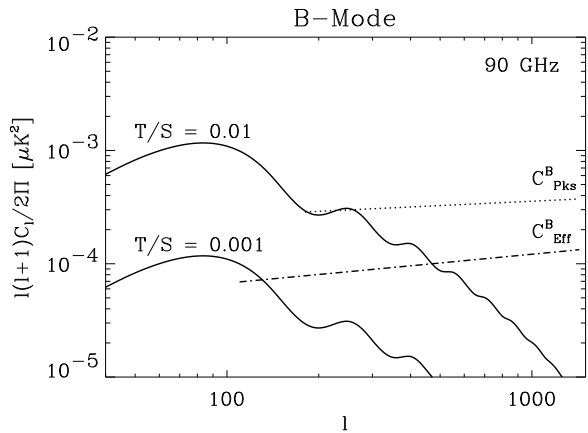}
\caption{$B$--mode angular power spectrum $C^B_{\rm Eff}$
of the 1.4-GHz data extrapolated to 90-GHz. 
A frequency spectral slope of $\alpha = -3.1$ is assumed.
The extrapolation of the 2.3-GHz spectrum obtained in the southern
area is also reported ($C^B_{\rm Pks}$).
Spectra expected for the CMBP are shown for comparison (solid) by using
tensor-to-scalar power ratios of $T/S = 0.01$ and $T/S = 0.001$, respectively.
The other cosmological parameters used are those of the concordance model
as determined in \citet{spergel03}.
\label{specCMB_B_Fig}}
\end{figure}
%%%%%%%%%%%%%%%%%%%%%%%%%%%%%%%%%%%%%%%%%%%%%%%%%%%%%%%%%%%%%%%%%%%%%%%%

Figure~\ref{specCMB_B_Fig} reports the $B$--mode spectrum extrapolated
to 90-GHz, that is in the frequency range expected to be the best
trade-off between Galactic synchrotron and dust contamination.
Even though the peak near $\ell$~$\sim$~$90$ (angular scale $\theta$~$\sim$~$2\degr$)
is marginally accessible by our data, the result is very promising.
The comparison with the expected CMBP signal for different $T/S$
values suggests that the properties of Inflation models with a GW amount
as low as $T/S = 0.001$ can be explored here, when only the synchrotron
contribution is considered. 
This result is even better than the one obtained in the southern region,
where $T/S$ values down to $\sim$~0.01 were estimated to be accessible.

As already pointed out by \citet{carretti05b}, this low level of the
polarized synchrotron emission can make the ISM dust the major
contaminant of CMBP at 90--100~GHz. 
No estimate of the polarized dust emission is available in our region,
and a real comparison is not possible. However, we can attempt a comparison
using the existing data.
The most proper one for our area would be the upper limit obtained by 
\citet{ponthieu05} with the data of 
the ARCHEOPS experiment, our area being included in the sky region they
surveyed. However, their result looks too pessimistic
to be applied to our low emission area. In fact, besides being an upper limit,
it refers to the whole $|b| > 10\degr$ portion of the region mapped by
that experiment. Moreover, the value they provided 
(0.2-$\mu$K$^2$ at 100-GHz) is much higher than the upper limit deduced
by \citet{carretti05b} relative to a large high Galactic latitude area.
Instead, we prefer to use the dust total intensity measurements
recently obtained at high Galactic latitudes
in the area surveyed with the 2003 BOOMERanG experiment \citep{masi05}.
They also find that the dust emission in that area 
is representative of a large fraction of the high latitude sky (40 per cent of the sky).
As polarization fraction we consider both 5 and 20~per~cent, which brackets
the 10~per~cent deduced by \citet{benoit04} for the high Galactic latitudes.

\begin{figure}
\centering
\includegraphics[angle=0, width=1.0\hsize]{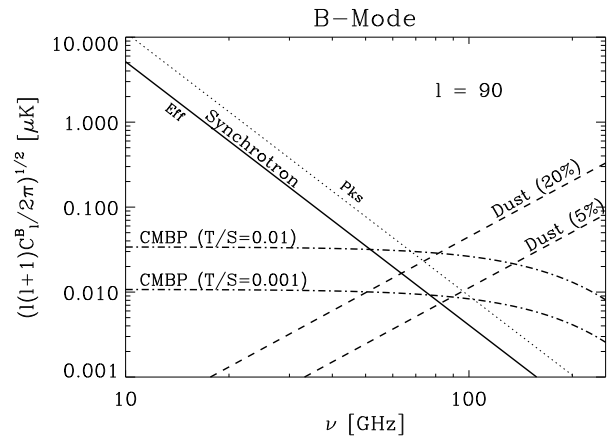}
\caption{Frequency behaviour of the $B$--mode for CMB and foregrounds
(synchrotron and dust). Emissions are estimated through the 
quantity $\sqrt{\ell (\ell + 1) C^B_l / (2 \pi)}$ 
at $\ell = 90$, since fair estimate of the signal
on the scale where the CMB $B$--mode peaks ($\theta$~$\sim$~$2^\circ$).
The expected CMB level is plotted in the case of both $T/S = 0.01$ and
$T/S = 0.001$.
Synchrotron emission is estimated by computing
the fit of Table~\ref{specTab} at $\ell = 90$ 
and assuming a slope of $\alpha = -3.1$ (solid).
The case of the southern area is shown as well for comparison (dotted).
Dust emission is evaluated considering the measurements of the  
total intensity signal performed with the 2003 BOOMERanG experiment 
in a high Galactic latitude region at 245-GHz 
and on an angular scale of 14-arcmin \citep{masi05}. A power spectrum 
behaviour of $C_\ell^{\rm dust} \propto \ell^{-2}$ \citep{bennett03} 
has been assumed to estimate the emission at $\ell = 90$.
A slope of $\alpha = 2.2$ is assumed for the frequency extrapolation 
\citep{bennett03}. Two values of polarization fraction (5 and 20~per~cent)
are used to bracket the 10~per~cent deduced by \citet{benoit04}
for high Galactic latitudes.
\label{foregFig}}
\end{figure}

\begin{figure}
\centering
\includegraphics[angle=0, width=1.0\hsize]{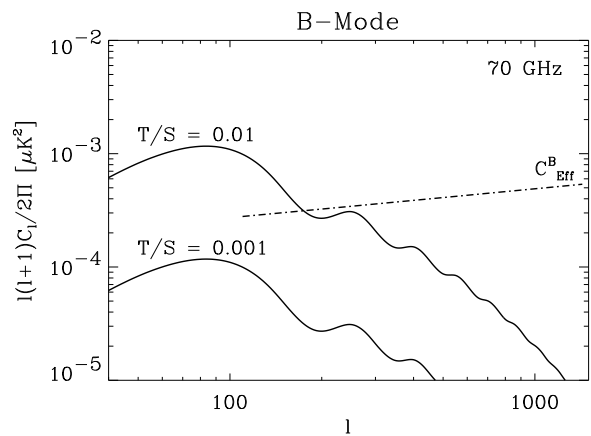}
\caption{As for Figure~\ref{specCMB_B_Fig}, 
but the power spectrum is extrapolated to 70-GHz. 
\label{specCMB_B_70_Fig}}
\end{figure}

The frequency behaviours are plotted in Figure~\ref{foregFig},
where the quantity $\sqrt{\ell(\ell+1)\,C^B_\ell / (2\pi)}$
at $\ell = 90$ is used as good indicator of the emission on
the $2\degr$ scale of the $B$--mode peak. 
The dust actually looks to be the leading
contaminant at 90--100~GHz, even assuming the case of 5~per~cent polarization. 
As a consequence, the best frequency window where the total foreground 
contamination is minimum can be {\it red}-shifted
toward $\sim70$~GHz, as it happens for the CMB anisotropy \citep{bennett03}.
To give an idea of the situation at 70-GHz, 
Figure~\ref{specCMB_B_70_Fig} shows the synchrotron angular
behaviour at such a frequency.

Besides implications for the design of experiments devoted to the CMBP
$B$-mode,  such a red-shift implies a reduction of the dust
contamination, because of its frequency behaviour. As a result, even
considering the dust contribution, it should be possible to study models
with $T/S = 0.01$, and even lower in case of 5~per~cent dust polarization.
As mentioned above, this result is even better than that obtained in the
other high Galactic latitude area, also reported in Figure~\ref{foregFig} 
for comparison.
Although observations in other sky regions are needed, 
this suggests that the southern area is not a special case, 
and that the possibility to reach the $T/S = 0.01$
value estimated to be achievable in the southern area
could be extended to large low emission 
regions at high Galactic latitudes.
Foreground cleaning techniques can allow further improvements of 
the measurable $T/S$ (e.g. \citealt{tegmark00,tucci05,verde05})
opening wide possibilities in the explorable Inflation models, especially 
in view of the recent results of \citet{boyle05}. 
In fact, 
these authors find that the interesting class of Inflation models
with minimal fine-tuning have $T/S > 0.01$.
Only models with an high degree of fine-tuning can have $T/S$ values 
less than 0.001. 
Thus, perspectives to detect the $B$-mode in large low emission areas 
at high Galactic latitudes become realistic.
However, a better assessment of the scenario requires
direct measures of the polarized dust emission, possibly at higher
frequency (hundreds of GHz) where this Galactic component is stronger.

\section*{Acknowledgments}

This work has been carried out in the framework of the BaR-SPOrt experiment, 
a programme funded by ASI (Agenzia Spaziale Italiana).
The paper is based on observations with the 100-m telescope of the MPIfR
(Max-Planck-Institut f\"ur Radioastronomie) at Effelsberg.
We like to thank an anonymous referee for useful comments, 
and O. Lochner for technical support with the 8-channel
polarization measurements. We acknowledge the use of the CMBFAST package.

%%%%%%%%%%%%%%%%%%%%%%%%%%%%%%%%%%%%%%%%%%%%%%%%%%%%%%%%%%%%%%%%%%%%%%%%
%%%%%%%%%%%%%%%%%%%%%%%%%%%%%%%%%%%%%%%%%%%%%%%%%%%%%%%%%%%%%%%%%%%%%%%%
%%%%%%%%%%%%%%%%%%%%%%%%%%%%%%%%%%%%%%%%%%%%%%%%%%%%%%%%%%%%%%%%%%%%%%%%
%%%%%%%%%%%%%%%%%%%%%%%%%%%%%%%%%%%%%%%%%%%%%%%%%%%%%%%%%%%%%%%%%%%%%%%%

\bsp

\label{lastpage}

\end{document}